\documentclass[longauth]{aa}

\usepackage{graphicx}

\usepackage[varg]{txfonts}

\usepackage{booktabs}
\usepackage{hyperref}
\usepackage{makecell}
\usepackage{orcidlink}

\newcommand{\hess}{H.E.S.S.\xspace}
\newcommand{\sgra}{Sgr~A$^\star$\xspace}
\newcommand{\hessGC}{HESS~J1745-290\xspace}
\newcommand{\hessPWN}{G$0.9+0.1$\xspace}
\newcommand{\GC}{GC\xspace}

\begin{document}
   
\title{Search for long-term variability of \hessGC}   

\author {
H.E.S.S. Collaboration : 
A.~Acharyya\inst{\ref{USD}} \orcidlink{0000-0002-2028-9230} % HESS member
\and F.~Aharonian\inst{\ref{DIAS},\ref{MPIK}}  \orcidlink{0000-0003-1157-3915} % HESS member
\and M.~Backes\inst{\ref{UNAM},\ref{NWU}}  \orcidlink{0000-0002-9326-6400} % HESS member
\and R.~Batzofin\inst{\ref{UP}}  \orcidlink{0000-0002-5797-3386} % HESS member
\and D.~Berge\inst{\ref{DESY},\ref{HUB}} \orcidlink{0000-0002-2918-1824} % HESS member
\and K.~Bernl\"ohr\inst{\ref{MPIK}}  \orcidlink{0000-0001-8065-3252} % HESS member
\and M.~B\"ottcher\inst{\ref{NWU}} \orcidlink{0000-0002-8434-5692} % HESS member
\and C.~Boisson\inst{\ref{LUX}}  \orcidlink{0000-0001-5893-1797} % HESS member
\and J.~Bolmont\inst{\ref{LPNHE}}  \orcidlink{0000-0003-4739-8389} % HESS member
\and F.~Brun\inst{\ref{IRFU}}  \orcidlink{0000-0003-0770-9007} % HESS member
\and B.~Bruno\inst{\ref{ECAP}}  % HESS member
\and C.~Burger-Scheidlin\inst{\ref{DIAS}}  \orcidlink{0000-0002-7239-2248} % HESS member
\and T.~Bylund\inst{\ref{LUX}}  % HESS member
\and J.~Celic\inst{\ref{ECAP}}  % HESS member
\and M.~Cerruti\inst{\ref{APC}} \orcidlink{0000-0001-7891-699X} % HESS member
\and A.~Chen\inst{\ref{Wits}}  \orcidlink{0000-0001-6425-5692} % HESS member
\and M.~Chernyakova\inst{\ref{DCU},\ref{DIAS}} \orcidlink{0000-0002-9735-3608} % HESS member
\and J. O.~Chibueze\inst{\ref{NWU},\ref{UNAM}}  \orcidlink{0000-0002-9875-7436} % HESS member
\and O.~Chibueze\inst{\ref{NWU}} \orcidlink{0000-0001-8601-2675} % HESS member
\and B.~Cornejo\inst{\ref{IRFU}}  \orcidlink{0009-0003-0039-0483} % HESS member
\and G.~Cotter\inst{\ref{UOX}}  \orcidlink{0000-0002-9975-1829} % HESS member
\and J.~Damascene~Mbarubucyeye\inst{\ref{DESY}}  \orcidlink{0000-0002-4991-6576} % HESS member
\and J.~de~Assis~Scarpin\inst{\ref{LLR}}  \orcidlink{0009-0004-4411-236X} % HESS member
\and M.~de~Bony~de~Lavergne\inst{\ref{IRFU},\ref{CPPM}}  \orcidlink{0000-0002-4650-1666} % HESS member
\and M.~de~Naurois\inst{\ref{LLR}}  \orcidlink{0000-0002-7245-201X} % HESS member
\and E.~de~O\~na~Wilhelmi\inst{\ref{DESY}}  \orcidlink{0000-0002-5401-0744} % HESS member
\and A.~G.~Delgado~Giler\inst{\ref{HUB}}  \orcidlink{0000-0003-2190-9857} % HESS member
\and J.~Devin\inst{\ref{LUPM}}  \orcidlink{0000-0003-1018-7246} % HESS member
\and A.~Djannati-Ata\"i\inst{\ref{APC}}  \orcidlink{0000-0002-4924-1708} % HESS member
\and A.~Dmytriiev\inst{\ref{NWU}}  \orcidlink{0000-0003-0102-5579} % HESS member
\and K.~Egg\inst{\ref{ECAP}}  \orcidlink{0009-0002-4238-034X} % HESS member
\and J.-P.~Ernenwein\inst{\ref{CPPM}}  % HESS member
\and C.~Esca\~{n}uela~Nieves\inst{\ref{MPIK}}  \orcidlink{0000-0002-7297-8126} % HESS member
\and P.~Fauverge\inst{\ref{LP2I}}  \orcidlink{0009-0006-1613-6633} % HESS member
\and K.~Feijen\inst{\ref{APC}}  \orcidlink{0000-0003-1476-3714} % HESS member
\and M.~D.~Filipovic\inst{\ref{Sydney}}  \orcidlink{0000-0002-4990-9288} % HESS member
\and G.~Fontaine\inst{\ref{LLR}}  \orcidlink{0000-0002-6443-5025} % HESS member
\and S.~Funk\inst{\ref{ECAP}}  \orcidlink{0000-0002-2012-0080} % HESS member
\and S.~Gabici\inst{\ref{APC}}  % HESS member
\and J.F.~Glicenstein\inst{\ref{IRFU}}  \orcidlink{0000-0003-2581-1742} % HESS member
\and J.~Glombitza\inst{\ref{ECAP}}  \orcidlink{0000-0001-9683-4568} % HESS member
\and P.~Goswami\inst{\ref{LSW}}  \orcidlink{0000-0001-5430-4374} % HESS member
\and L.~Heckmann\inst{\ref{APC}}  \orcidlink{0000-0002-6653-8407} % HESS member
\and B.~Hess\inst{\ref{IAAT}}  \orcidlink{0009-0004-9999-171X} % HESS member
\and J.A.~Hinton\inst{\ref{MPIK}}  \orcidlink{0000-0002-1031-7760} % HESS member
\and W.~Hofmann\inst{\ref{MPIK}}  \orcidlink{0000-0001-8295-0648} % HESS member
\and T.~L.~Holch\inst{\ref{DESY}}  \orcidlink{0000-0001-5161-1168} % HESS member
\and M.~Holler\inst{\ref{Innsbruck}}  \orcidlink{0000-0002-0107-8657} % HESS member
\and D.~Horns\inst{\ref{UHAM}}  \orcidlink{0000-0003-1945-0119} % HESS member
\and M.~Jamrozy\inst{\ref{OAUJ}}  \orcidlink{0000-0002-0870-7778} % HESS member
\and F.~Jankowsky\inst{\ref{LSW}}  % HESS member
\and I.~Jaroschewski\inst{\ref{IRFU}}  \orcidlink{0000-0001-5180-2845} % HESS member
\and I.~Jung-Richardt\inst{\ref{ECAP}}  % HESS member
\and D.~Kerszberg\inst{\ref{LPNHE}}  \orcidlink{0000-0002-5289-1509} % HESS member
\and B. Khélifi\inst{\ref{APC}}  \orcidlink{0000-0001-6876-5577} % HESS member
\and N.~Komin\inst{\ref{LUPM},\ref{Wits}} \orcidlink{0000-0003-3280-0582} % HESS member
\and K.~Kosack\inst{\ref{IRFU}} \orcidlink{0000-0001-8424-3621} % HESS member
\and D.~Kostunin\inst{\ref{DESY}} \orcidlink{0000-0002-0487-0076} % HESS member
\and R.G.~Lang\inst{\ref{ECAP}} \orcidlink{0000-0003-0492-5628} % HESS member
\and S.~Lazarevi\'c\inst{\ref{Sydney}}  \orcidlink{0000-0001-6109-8548} % HESS member
\and A.~Lemi\`ere\inst{\ref{APC}} \textsuperscript{*} \orcidlink{0000-0002-6682-7188} % HESS member
\and M.~Lemoine-Goumard\inst{\ref{LP2I}}  \orcidlink{0000-0002-4462-3686} % HESS member
\and J.-P.~Lenain\inst{\ref{LPNHE}}  \orcidlink{0000-0001-7284-9220} % HESS member
\and P.~Liniewicz\inst{\ref{OAUJ}} \orcidlink{0009-0008-3575-3965} % HESS member
\and J.~Mackey\inst{\ref{DIAS}}\orcidlink{0000-0002-5449-6131} % HESS member
\and D.~Malyshev\inst{\ref{IAAT}} \orcidlink{0000-0001-9689-2194} % HESS member
\and V.~Marandon\inst{\ref{IRFU}} \orcidlink{0000-0001-9077-4058} % HESS member
\and M.~G.~F.~Mayer\inst{\ref{ECAP}} \orcidlink{0000-0002-9771-9841} % HESS member
\and A.~Mehta\inst{\ref{DESY}} % HESS member
\and A.M.W.~Mitchell\inst{\ref{ECAP}} \orcidlink{0000-0003-3631-5648} % HESS member
\and R.~Moderski\inst{\ref{NCAC}} \orcidlink{0000-0002-8663-3882} % HESS member
\and L.~Mohrmann\inst{\ref{MPIK}}  \orcidlink{0000-0002-9667-8654} % HESS member
\and A.~Montanari\inst{\ref{LSW}}  \orcidlink{0000-0002-3620-0173} % HESS member
\and J.~Niemiec\inst{\ref{IFJPAN}}  \orcidlink{0000-0001-6036-8569} % HESS member
\and L.~Olivera-Nieto\inst{\ref{GRAPPA}}  \orcidlink{0000-0002-9105-0518} % HESS member
\and M.O.~Moghadam\inst{\ref{UP}}  \orcidlink{0009-0003-2479-1863} % HESS member
\and S.~Panny\inst{\ref{Innsbruck}}  \orcidlink{0000-0001-5770-3805} % HESS member
\and R.D.~Parsons\inst{\ref{HUB}}  \orcidlink{0000-0003-3457-9308} % HESS member
\and U.~Pensec\inst{\ref{LPNHE}}  \orcidlink{0009-0009-2359-1775} % HESS member
\and P.~Pichard\inst{\ref{APC}}  \orcidlink{0009-0005-9803-0762} % HESS member
\and T.~Preis\inst{\ref{Innsbruck}}  \orcidlink{0009-0001-7110-6764} % HESS member
\and G.~P\"uhlhofer\inst{\ref{IAAT}}  \orcidlink{0000-0003-4632-4644} % HESS member
\and M.~Punch\inst{\ref{APC}}  \orcidlink{0000-0002-4710-2165} % HESS member
\and A.~Quirrenbach\inst{\ref{LSW}}  % HESS member
\and A.~Reimer\inst{\ref{Innsbruck}}  \orcidlink{0000-0001-8604-7077} % HESS member
\and O.~Reimer\inst{\ref{Innsbruck}}  \orcidlink{0000-0001-6953-1385} % HESS member
\and I.~Reis\inst{\ref{IRFU}}  % HESS member
\and H.~X.~Ren\inst{\ref{MPIK}}  \orcidlink{0000-0003-0221-2560} % HESS member
\and B.~Reville\inst{\ref{MPIK}}  \orcidlink{0000-0002-3778-1432} % HESS member
\and F.~Rieger\inst{\ref{MPIK}}  % HESS member
\and G.~Rowell\inst{\ref{Adelaide}}  \orcidlink{0000-0002-9516-1581} % HESS member
\and B.~Rudak\inst{\ref{NCAC}}  \orcidlink{0000-0003-0452-3805} % HESS member
\and K.~Sabri\inst{\ref{LUPM}}  % HESS member
\and V.~Sahakian\inst{\ref{YPI}}  \orcidlink{0000-0003-1198-0043} % HESS member
\and A.~Santangelo\inst{\ref{IAAT}}  \orcidlink{0000-0003-4187-9560} % HESS member
\and M.~Sasaki\inst{\ref{ECAP}}  \orcidlink{0000-0001-5302-1866} % HESS member
\and F.~Sch\"ussler\inst{\ref{IRFU}}  \orcidlink{0000-0003-1500-6571} % HESS member
\and W.~Si~Said\inst{\ref{LLR}}  \orcidlink{0009-0007-6555-6893} % HESS member
\and H.~Sol\inst{\ref{LUX}}  % HESS member
\and {\L.}~Stawarz\inst{\ref{OAUJ}}  \orcidlink{0000-0002-7263-7540} % HESS member
\and T.~Tanaka\inst{\ref{Konan}}  \orcidlink{0000-0002-4383-0368} % HESS member
\and G.~L.~Taylor\inst{\ref{LSW}}  \orcidlink{0009-0001-8062-036X} % HESS member
\and R.~Terrier\inst{\ref{APC}} \textsuperscript{*} \orcidlink{0000-0002-8219-4667} % HESS member
\and M.~Tsirou\inst{\ref{DESY}}  \orcidlink{0000-0003-3417-1425} % HESS member
\and T.~Unbehaun\inst{\ref{ECAP}}  \orcidlink{0000-0002-7378-4024} % HESS member
\and C.~van~Eldik\inst{\ref{ECAP}}  \orcidlink{0000-0001-9669-645X} % HESS member
\and M.~Vecchi\inst{\ref{Groningen}}  \orcidlink{0000-0002-5338-6029} % HESS member
\and C.~Venter\inst{\ref{NWU}}  \orcidlink{0000-0002-2666-4812} % HESS member
\and J.~Vink\inst{\ref{GRAPPA}}  \orcidlink{0000-0002-4708-4219} % HESS member
\and V.~Voitsekhovskyi\inst{\ref{GRAPPA}}  \orcidlink{0000-0002-3906-4840} % HESS member
\and T.~Wach\inst{\ref{ECAP}}  \orcidlink{0009-0008-4658-7405} % HESS member
\and S.J.~Wagner\inst{\ref{LSW}}  \orcidlink{0000-0002-7474-6062} % HESS member
\and A.~Wierzcholska\inst{\ref{IFJPAN},\ref{LSW}}  \orcidlink{0000-0003-4472-7204} % HESS member
\and M.~Zacharias\inst{\ref{LSW},\ref{NWU}}  \orcidlink{0000-0001-5801-3945} % HESS member
\and A.~Zech\inst{\ref{LUX}}  % HESS member
\and W.~Zhong\inst{\ref{DESY}}  \orcidlink{0000-0003-3717-2861} % HESS member
\and S.~Zouari \textsuperscript{*}\inst{\ref{APC}}
}

\institute{
\raggedright
University of Southern Denmark \label{USD}
\and Astronomy \& Astrophysics Section, School of Cosmic Physics, Dublin Institute for Advanced Studies, DIAS Dunsink Observatory, Dublin D15 XR2R, Ireland \label{DIAS}
\and Max-Planck-Institut für Kernphysik, P.O. Box 103980, D 69029 Heidelberg, Germany \label{MPIK}
\and University of Namibia, Department of Physics, Private Bag 13301, Windhoek 10005, Namibia \label{UNAM}
\and Centre for Space Research, North-West University, Potchefstroom 2520, South Africa \label{NWU}
\and Institut für Physik und Astronomie, Universität Potsdam, Karl-Liebknecht-Strasse 24/25, D 14476 Potsdam, Germany \label{UP}
\and Deutsches Elektronen-Synchrotron DESY, Platanenallee 6, 15738 Zeuthen, Germany \label{DESY}
\and Institut für Physik, Humboldt-Universität zu Berlin, Newtonstr. 15, D 12489 Berlin, Germany \label{HUB}
\and LUX, Observatoire de Paris, Université PSL, CNRS, Sorbonne Université, 5 Pl. Jules Janssen, 92190 Meudon, France \label{LUX}
\and Sorbonne Université, CNRS/IN2P3, Laboratoire de Physique Nucléaire, et de Hautes Energies, LPNHE, 4 place Jussieu, 75005 Paris, France \label{LPNHE}
\and IRFU, CEA, Université Paris-Saclay, F-91191 Gif-sur-Yvette, France \label{IRFU}
\and Friedrich-Alexander-Universität Erlangen-Nürnberg, Erlangen Centre for Astroparticle Physics,  Nikolaus-Fiebiger-Str. 2, 91058 Erlangen, Germany \label{ECAP}
\and Université Paris Cité, CNRS, Astroparticule et Cosmologie, F-75013 Paris, France \label{APC}
\and School of Physics, University of the Witwatersrand, 1 Jan Smuts Avenue, Braamfontein, Johannesburg, 2050, South Africa \label{Wits}
\and School of Physical Sciences and Centre for Astrophysics \& Relativity, Dublin City University, Glasnevin, Dublin D09 W6Y4, Ireland \label{DCU}
\and University of Oxford, Department of Physics, Denys Wilkinson Building, Keble Road, Oxford OX1 3RH, UK, United Kingdom \label{UOX}
\and Laboratoire Leprince-Ringuet, École Polytechnique, CNRS, Institut Polytechnique de Paris, F-91128 Palaiseau, France \label{LLR}
\and Aix Marseille Université, CNRS/IN2P3, CPPM, Marseille, France \label{CPPM}
\and Laboratoire Univers et Particules de Montpellier, Université Montpellier, CNRS/IN2P3, CC 72, Place Eugène Bataillon, F-34095 Montpellier Cedex 5, France \label{LUPM}
\and Université Bordeaux, CNRS, LP2I Bordeaux, UMR 5797, F-33170 Gradignan, France \label{LP2I}
\and School of Science, Western Sydney University, Locked Bag 1797, Penrith South DC, NSW 2751, Australia \label{Sydney}
\and Landessternwarte, Universität Heidelberg, Königstuhl, D 69117 Heidelberg, Germany \label{LSW}
\and Institut für Astronomie und Astrophysik, Universität Tübingen, Sand 1, D 72076 Tübingen, Germany \label{IAAT}
\and Universität Innsbruck, Institut für Astro- und Teilchenphysik, Technikerstraße 25, 6020 Innsbruck, Austria \label{Innsbruck}
\and Universität Hamburg, Institut für Experimentalphysik, Luruper Chaussee 149, D 22761 Hamburg, Germany \label{UHAM}
\and Obserwatorium Astronomiczne, Uniwersytet Jagielloński, ul. Orla 171, 30-244 Kraków, Poland \label{OAUJ}
\and Nicolaus Copernicus Astronomical Center, Polish Academy of Sciences, ul. Bartycka 18, 00-716 Warsaw, Poland \label{NCAC}
\and Instytut Fizyki Jac{a}drowej PAN, ul. Radzikowskiego 152, ul. Radzikowskiego 152, 31-342 Kraków, Poland \label{IFJPAN}
\and GRAPPA, Anton Pannekoek Institute for Astronomy, University of Amsterdam, Science Park 904, 1098 XH Amsterdam, The Netherlands \label{GRAPPA}
\and School of Physical Sciences, University of Adelaide, Adelaide 5005, Australia \label{Adelaide}
\and Yerevan Physics Institute, 2 Alikhanian Brothers St., 0036 Yerevan, Armenia \label{YPI}
\and Department of Physics, Konan University, 8-9-1 Okamoto, Higashinada, Kobe, Hyogo 658-8501, Japan \label{Konan}
\and Kapteyn Astronomical Institute, University of Groningen, Landleven 12, 9747 AD Groningen, The Netherlands \label{Groningen}
}

\offprints{\texttt{contact.hess@hess-experiment.eu}\\ 
\textsuperscript{*} Corresponding authors}

\abstract
    {
    At the center of our Galaxy lies the bright $\gamma$-ray point-like source \hessGC, which is compatible in position with \sgra, although an association between the two remains uncertain. Using data obtained between 2004 and 2019 with the High Energy Stereoscopic System (\hess) on the Galactic center region, we studied the variability of \hessGC over 353 hours of observations collected over 16 years, representing the largest dataset gathered yet on this region at TeV energies. We performed a 3D maximum-likelihood analysis of the central source and the diffuse $\gamma$-ray emission in the Galactic center region. This analysis allowed us to extract the spectral and morphological intrinsic behavior of the two components. 
    By performing this analysis on an annual basis, we derived the light curve of \hessGC and the diffuse emission over the past 16 years. The 3D maximum-likelihood analysis method allowed us to separate the central source from the overlapping diffuse emission, enabling a recalibration of the former by the latter and alleviating some of the systematic effects. We find no long-term or yearly variability. We also provide an estimate of the sensitivity of \hess to variation of this specific source over 16 years. We rule out any yearly $\gamma$-ray flux variation of this source larger than 30\%, as well as any linear flux variation exceeding 30\% over this time period.
    }

    \keywords{Galaxy: center -- gamma rays: general -- gamma rays: ISM
                  }

\date{Received 13 October 2025 / Accepted for publication 14 February 2026}

%\authorrunning{H.E.S.S. Collaboration}
\maketitle

%\makeatletter
%\renewcommand*{\@fnsymbol}[1]{\ifcase#1\@arabic{#1}\fi}
%\makeatother

\section{Introduction}\label{section1}
The Galactic center (GC) is a prime observation region for very high energy (VHE) $\gamma$ rays. Various observatories, including the High Energy Stereoscopic System (\hess) \citep{2004AetA...425L..13A,2006PhRvL..97v1102A,2008AetA...492L..25A,2009AetA...503..817A, 2010MNRAS.402.1877A}, the Major Atmospheric Gamma-ray Imaging Cherenkov Telescope (MAGIC) \citep{2006ApJ...638L.101A, 2020AetA...642A.190M}, and the Very Energetic Radiation Imaging Telescope Array System (VERITAS) \citep{2014ApJ...790..149A, 2016ApJ...821..129A, 2021ApJ...913..115A} have detected the point-like source \hessGC. 
This source is centered at the Galactic coordinates $l=359.944^\circ$, $b=-0.046^\circ$, with a 13 arcsec uncertainty and an upper bound on its extension of about 1.2\,arcmin \citep{2010MNRAS.402.1877A}.

Although \hessGC is the most prominent feature of the region, \hess has also identified diffuse emission extending over 200 pc along the inner Galactic plane\citep{2006Natur.439..695A, 2018AetA...612A...9H}. 
This diffuse emission most likely indicates a high density of cosmic rays in the central region of the Galaxy \citep{2016Natur.531..476H}, which interact with the central molecular zone (CMZ) to produce $\gamma$ rays via proton-proton interaction and neutral pion decay.

Despite numerous studies, the physical origin of \hessGC remains an open question.
Potential sources such as the supernova remnant Sgr\,A East have been ruled out based on their extension and positional incompatibility with the source  \citep{2010MNRAS.402.1877A}. The pulsar wind nebula (PWN) candidate G359.95-0.04 remains a plausible counterpart, as it is compatible in both position and energetics with the source \citep{2006MNRAS.367..937W, 2007ApJ...657..302H}. However, the estimated magnetic field in the vicinity of \GC might be too large for this scenario to be plausible \citep{2015arXiv151101159K}.
The super-massive black hole (SMBH) at the center of our Galaxy, \sgra,  is therefore a plausible origin of the $\gamma$-ray source, as it is fully compatible in position and its accretion flow provides a site of particle acceleration.

Across the electromagnetic spectrum, \sgra shows high and rapid variability, with the highest flux variations observed in the X-ray and infrared (IR) domain (see overviews \cite{2010RvMP...82.3121G,2021ApJ...917...73W}).
X-ray observatories have observed regular flares that last a few hours and occur daily. These events have varying intensities that can reach up to two orders of magnitude above the quiescent level (for Chandra, see \cite{2001Natur.413...45B,2015ApJ...799..199N}; for Swift, see \cite{2022MNRAS.510.2851A}).
In the IR domain, the emission of \sgra shows high variability \citep{2003Natur.425..934G}; however, additional individual flaring events have recently been identified in the NIR range, with  amplitudes reaching a factor of two on timescales of $\sim 20$ minutes \citep{2019ApJ...882L..27D,2020AetA...638A...2G}. 

The link between IR and X-ray flares is not completely straightforward, as X-ray flares are usually accompanied by a flare-type event in IR \citep{2004AetA...427....1E}, but the reverse is not always true. The emission mechanism behind these flares is generally interpreted as lepton-based synchrotron and/or Compton-scattering emission occurring in the inner accretion flow, relatively close to the SMBH (from a few Schwarzschild radii, $R_S$, up to $\sim 10^2 R_S$) \citep{2003ApJ...598..301Y, 2010ApJ...725..450D,2017MNRAS.468.2447P, 2021AetA...654A..22G}.

The high-energy particles that produce the $\gamma$-ray signal can be accelerated close to the SMBH, but they can also escape the accretion flow and interact farther in the surrounding medium, depending on the particle transport conditions and the matter distribution. For instance, \citet{2006ApJ...647.1099L} proposed that protons are accelerated by stochastic processes within the inner 20 $R_S$ and escape and scatter in the inner few parsecs to produce the TeV emission.  As noted by \citet{2007ApJ...657L..13B} and \citet{2012ApJ...753...41L} , the most plausible target for producing $\gamma$ rays is the circumnuclear disk (CND) \citep{2005ApJ...622..346C}, a 1.5 pc radius ring-like structure of dense molecular matter surrounding \sgra. Depending on the assumed particle transport hypotheses, the GC VHE source could therefore reflect the history of particle acceleration in the vicinity of the SMBH over shorter or longer timescales and remain stable on timescales of years or more \citep{2011MNRAS.410.1521B, 2011ApJ...726...60C, 2013JPhG...40f5201G}. The long-term variability of \hessGC is therefore a key observable to determine its nature and possibly constrain the particle acceleration processes in \sgra. 
Whipple first studied the variability of the \GC at VHE between 2000 and 2004 \citep{whipple2004} and showed that the source's flux was consistent with being constant. Another study in 2009 \citep{2009AetA...503..817A} used the first three years of \hess data (2004-2006) to search for variability on 28-minute timescales, hour-long flares, and quasi-periodic oscillations (QPOs), but detected no significant signals. 
More recent studies  by MAGIC \citep{2017AetA...601A..33A} and VERITAS \citep{2014ApJ...790..149A,2021ApJ...913..115A}, corroborate this apparent stability over daily to monthly timescales. 
In addition, attempts to detect counterparts of X-ray flares with \hess have so far been unsuccessful (see \cite{2008AetA...492L..25A} for the joint \hess-Chandra observation campaign in 2006).

In this work, we investigate the potential variability of \hessGC over timescales of years. We take advantage of the large \hess observation database to measure the yearly evolution of the VHE flux over the 16-year period from 2004 to 2019, driven by the parsec-scale $\gamma$-ray emission implied by models.

After describing how we selected and processed the 16 years of \hess data on the \GC (Section \ref{section2}), we present our methodology in two parts.
First, we describe the spectro-morphological analysis of \hessGC and its surroundings (Section \ref{section3}), then the study of time variability of \hessGC (Section \ref{section4}). 
In Section \ref{section5}, we reassess the sensitivity of the \hess survey to several theoretical time variation models.
Finally, in Section \ref{section6}, we discuss the significance of these results in the context of existing knowledge of \sgra variability across the electromagnetic spectrum.

\section{Data reduction}\label{section2}

The \hess array consists of five imaging atmospheric cherenkov telescopes (IACTs), located in Namibia, and has been observing the southern hemisphere sky at VHE since 2003.
Observations by \hess have monitored the GC ever since 
(albeit not in a uniform manner, as shown in the next section), making it one of the most frequently observed regions of the TeV sky.

\subsection{Data selection and processing}\label{section21}

In this study, we considered all \hess data available for the \GC region between 2004 and 2019. 
To ensure a homogeneous and robust long-term variability analysis, we restricted our dataset to Galactic center observations acquired before 2020, prior to important changes in the exposure pattern and the upgrade of the camera and data-acquisition systems.
We chose observation ``runs'' (uninterrupted observation segments of typically 28 minutes) pointing within 1.8 degrees of the position of \sgra. 
We set the maximum zenith angle to 50 degrees and used only runs involving all four 12 m-diameter medium-size telescopes (CT1-4), which have been operational since 2003. 
This selection results in a total of 768 runs and 
353 hours of observation time after standard quality cuts \citep{2015ICRC...34..837K}.  Table \ref{tab:runs} summarizes the observation properties per year.

\begin{table}
    \centering
    \caption{Observation summary by year.}
    \begin{tabular}{ccccc}
    \toprule
    \toprule
         \textbf{year} & \textbf{runs} & \textbf{livetime (h)} & \makecell{\textbf{avg zenith} \\\textbf{angle ($^\circ$)}} & \makecell{\textbf{average optical} \\\textbf{efficiency}} \\
         \midrule
2004 & 99 & 44.6 & 20.7 & 0.80 \\
2005 &  116 & 53.0 & 22.7 & 0.68 \\
2006 & 36 & 16.6 & 18.7 & 0.63 \\
2007 & 15 & 7.0 & 11.1 & 0.63 \\
2008 & 24 & 11.0 & 15.4 & 0.58 \\
2009 & 7 & 3.2 & 15.5 & 0.57 \\
2010 & 19 & 12.1 & 11.4 & 0.63 \\
2011 & 19 & 8.6 & 32.3 & 0.63 \\
2012 & 84 & 37.4 & 19.2 & 0.66 \\
2013 & 105 & 47.2 & 22.8 & 0.66 \\
2014 & 120 & 53.6 & 19.7 & 0.63 \\
2015 & 57 & 24.8 & 17.3 & 0.63 \\
2016 & 11 & 5.0 & 12.5 & 0.59 \\
2017 & 2 & 0.8 & 25.5 & 0.85 \\
2018 & 14 & 6.3 & 21.6 & 0.82 \\
2019 & 49 & 21.4 & 33.4 & 0.77 \\
\midrule
\textbf{Dataset} & 777& 352.7 & 20.6 & 0.68 \\

         \bottomrule
    \end{tabular}
    \tablefoot{% 
    Number of runs after run quality selection, total live-time, average zenith angle, and average optical efficiency measured from muon images\protect\footnotemark, for each year and for the full dataset. 
    }
    \label{tab:runs}
\end{table}

\footnotetext{This optical efficiency is derived from the observation of muon events with \hess. Comparing this measurement to the expected amount of Cherenkov light detected provides an estimate of the optical throughput of the instrument (between 0 and 1)\citep{chalmecalvet2014muonefficiencyhesstelescope}.}

We used a list of $\gamma$-ray candidates obtained from the HAP-fr analysis pipeline \citep{2015ICRC...34..837K} which reconstructs event showers based on the Hillas reconstruction \citep{1985ICRC....3..445H} and discriminates between $\gamma$ rays and hadronic events with the multivariate analysis (MVA) technique \citep{2012AIPC.1505..741B}.  
Using a configuration optimized for Galactic sources (i.e., weak, extended, and hard sources) allowed us to include data down to 300\,GeV. 
We then exported the output of the data processing to a FITS format developed by \cite{2017AIPC.1792g0006D}, which includes the list of $\gamma$-ray candidates, the instrument response functions (IRFs) (effective area, energy dispersion, exposure live time, and point spread function), as well as a residual hadronic background model.
Next, we used Gammapy 
\citep{gammapy:2023}, an open source Python library, to perform the high-level analysis, produce spectra and light curves, and perform simulations to assess systematic uncertainties.

We thoroughly cross-checked all results presented in this work using data analyzed by an alternative analysis pipeline: the HAP standard ImPACT method \citep{2009APh....31..383O,parsons_hinton2014}, and applying the same criteria for run selection, the same spatial offset requirement, and the same method to determine the energy thresholds. 
We performed high-level analysis on each dataset using the same Gammapy-based procedure.

\subsection{Data cube creation} 

The next step of the analysis consisted of converting the list of candidates for $\gamma$-rays into a data cube, i.e., a 3D array counting events by direction (two spatial dimensions) and by energy (one dimension). This data cube comes with the IRFs and the background model. The technique of fitting both spatial and spectral components simultaneously is hereafter referred to as the 3D analysis. 
In our analysis, we projected $\gamma$-ray candidates into a binned data cube over a spatial region of $4^\circ \times 3^\circ$, centered on the \GC, with pixels of size $0.02^\circ \times 0.02^\circ$. 
We rejected events with offsets above 1.8$^\circ$ from the center of the camera to avoid poorly reconstructed $\gamma$ events.
We divided the energy binning of the data cube into 25 logarithmic bins, covering 300\,GeV to 50\,TeV. 
For each run, we applied a safe energy threshold at the energy where the effective area drops to 10\% of its maximum value. Similarly to the procedure followed by \citet{2021AetA...653A.152A}, we imposed an additional energy threshold during the analysis, limiting the study to energy ranges where the background model (see Section 2.3) adequately describes the cosmic-ray background \citep{2019AetA...632A..72M}.
 This procedure results in most runs having a threshold around 400\,GeV. 
We produced a data cube for each observation and then stacked them into a global data cube for each year, which involved weighted averaging over the IRFs to obtain a single set of IRFs for each final dataset.

\subsection{Background template creation}

To account for the background in the 3D analysis, we evaluated the background level in every bin of the data cube. We derived the background model by projecting observations taken at high Galactic latitudes into a multidimensional table as a function of observation conditions, while masking known $\gamma$-ray sources.
To obtain a background model prediction for each run, we applied a multivariable interpolation based on observation conditions. This process is described in detail in \cite{2019AetA...632A..72M} and \cite{2021AetA...653A.152A}.
The resulting run-wise background model is a function of the measured energy and the reconstructed position in the field of view. Because of the large normalization uncertainties (see, e.g.,\cite{2019AetA...632A..72M}), we normalized run-wise models on each observation data cube, excluding $\gamma$-ray bright regions, following the field-of-view method described in \cite{2007AetA...466.1219B}.
We set an exclusion region extending over $3^\circ$ in longitude and $1^\circ$ in latitude, 
centered on the \GC, as well as a disk-shaped region of $0.7^\circ$ radius masking HESS$~$J1745-303. We then applied this procedure to each observation and stacked the resulting templates. 
To account for possible systematic uncertainties in the template normalization, we included it as a nuisance parameter when fitting a sky model to the observed data.

\section{Total dataset spectro-morphological analysis}\label{section3} 
Deriving a long-term light curve of the point source \hessGC requires describing both \hessGC and the diffuse ridge emission that covers the central few degrees, simultaneously and independently, as well as other compact sources such as \hessPWN \citep{2005AetA...432L..25A} or HESS$~$J1746-285 \citep{2018AetA...612A...9H}. Standard 1D spectral analysis methods cannot differentiate between two superimposed sources, which complicates studies of the GC region.
The 3D analysis allowed us to fit a spectro-morphological model to a data cube. 
Its ability to separate different physical components has already been demonstrated in VHE \citep{2021AetA...653A.152A}\citep{2023AetA...672A.103H}.
We used here the binned version of the 3D analysis as implemented in Gammapy and validated by previous studies \citep{2019AetA...632A..72M}.
This method estimates physical model parameters by comparing the cube of measured counts in space and energy with the model prediction, convolved with the IRFs, and including a model of the residual hadronic emission.
We performed this comparison by maximizing a likelihood function using the ``Cash'' statistic \citep{1979ApJ...228..939C}.

\subsection{Source model description}\label{section31}

We assume that the spatial parameters and most spectral parameters (except all normalizations) in our model do not vary with time.
We first performed a spectro-morphological fit on the complete dataset.
The model includes \hessGC, the diffuse emission $($DE$)$
, and two point sources: G\,0.9+0.1 and HESS\,J1746-285. 
The base source model hypotheses are similar to those of the \hess diffuse emission study \citep{2018AetA...612A...9H}. We use a spectral description for \hessGC and the DE as in \citep{2016Natur.531..476H}: a power-law with exponential cutoff for the point source, and a simple power-law for the DE. 
The spectra of G\,0.9+0.1 and HESS\,J1746-285 follow power laws, as described in \cite{2018AetA...612A...9H}.
A detailed spectral description of the DE is beyond the scope of this article. Our spatial description uses a spatial template similar to that in \cite{2018AetA...612A...9H}, based on a 2D distribution of cosmic rays and a velocity-integrated map of the CS emission in the CMZ \citep{tsuboi1999} for the DE and including a 2D template of the large-scale unresolved Galactic emission. 
We refit spectral parameters, source positions, and morphological template parameters to accurately reproduce the extra data collected and the new 3D description.

\subsection{Model fitting and spectral extraction}\label{section32}
We left all spectral parameters of \hessGC and the DE free (normalization, spectral index, and cut-off energy, when applicable), as well as the position of the point source.
The fitted spectral parameters for \hessGC are 
$\Phi = (2.32 \pm 0.06_{stat}\;^{+0.57}_{-0.42}\,_{syst}) \times 10^{-12} \rm cm^{-2}\rm s^{-1}\rm TeV^{-1}$, 
$\Gamma = (1.94 \pm 0.03_{stat} \pm 0.10_{syst}$), and $E_{cut} = (8.8 \pm 1.1_{stat} \;^{+1.73}_{-1.24}\,_{syst}) \rm TeV $. 
We evaluated systematic errors in the spectral parameters of \hessGC, using Monte Carlo simulations performed with Gammapy. These simulations included the uncertainties on the IRFs and background modeling, since they are often considered the dominant contributors. These are linked to variations in the atmospheric conditions, the presence of broken pixels in the camera, and the uncertainty in the absolute calibration of the telescopes \citep{Aharonian2006:Crab}. 
Taking systematic errors into account, the derived spectrum is compatible with previous \hess estimates for the \hessGC spectrum \citep{2016Natur.531..476H}, as well as the latest MAGIC \citep{2020AetA...642A.190M} and VERITAS \citep{2021ApJ...913..115A} measurements.
However, it is important to remember that these previous measurements were not intrinsic spectra, as they also incorporated the contribution of the DE within a region of $0.1^\circ$ around \sgra, and 
therefore cannot be expected to produce parameters identical to our measurement. 

\section{Variability study}\label{section4}
\subsection{General method}\label{section41}

To measure the evolution of the flux of \hessGC over 16 years of monitoring, we constructed 3D datasets of the region on a yearly time scale. 
Table \ref{tab:runs} lists the resulting total exposure per yearly dataset.
We then refit each dataset using the model previously applied in the 3D analysis of the total dataset (see Section \ref{section3}). In this process, we fixed all parameters from the total dataset-fit , except for the fluxes of \hessGC and the DE, for which we left the normalizations free to vary.
In this temporal analysis, we faced the challenge of fitting the model across a heterogeneous set of datasets. Some years have extensive hours of observation, while others have very limited data. Additionally, observation conditions vary from year to year, leading to changes in the energy threshold over time. To ensure that the flux measurements for DE and \hessGC were comparable across all years, we reduced the energy interval used in the model fit to 650\,GeV - 50\,TeV. This reduction in the energy range ensured consistency, allowing us to compare spectra 
more reliably across all datasets. We then constructed light curves using yearly flux measurements computed over the 1–10 TeV energy range.

By leaving the normalization of the DE free during the fit, we expected to see some variations of its flux (see results section 4.2). An important part of these variations originates from statistical fluctuations, which was accounted for by the statistical error bars, but residual fluctuations associated with systematic uncertainties remain. 
Indeed, the degradation and modifications of the instrument over time, as well as changing weather conditions, can influence annual monitoring. Hence, even after correcting for the varying optical efficiency, the IRFs of observations performed under different instrumental conditions are likely to contain residual systematic errors, which can lead to unstable measurements of the flux of a steady source. 
Consequently, to study the variability of \hessGC, time-dependent systematic effects must be taken into account. 
A simple way to alleviate time-dependent systematic effects is to correct the time-dependent IRF uncertainties using another source in the field of view that is expected to be non-variable. As its flux is comparable to that of the \GC source and because it must be constant over the 16 years covered, the DE is a good candidate for the recalibration source. The recalibration process implicitly uses the DE fluctuations as a yearly estimate of the instrumental deviation to correct the  \hessGC light curve. 
For each year, we applied the following correction to the source of interest \hessGC:
$$ F_{GC, recal}(t) =  \frac{F_{GC}(t)}{F_{DE}(t)} \times \overline{F_{DE}}, $$
where $F_{GC}$ and $F_{DE}$ the fluxes of \hessGC and the DE measured at a given time,  $\overline{F_{DE}}$ is the reference flux of the DE (weighted average), and $F_{GC, recal}$ is the recalibrated flux of \hessGC.
This is equivalent to studying the relative flux of \hessGC with respect to the DE.

This process implies an increase in the statistical fluctuations of the corrected \hessGC light curve, but it ensures that the light curve is corrected for any residual underestimated instrumental drift over time.

\subsection{Results}\label{section42}
\begin{figure}
    \centering
    \includegraphics[width=0.48\textwidth]{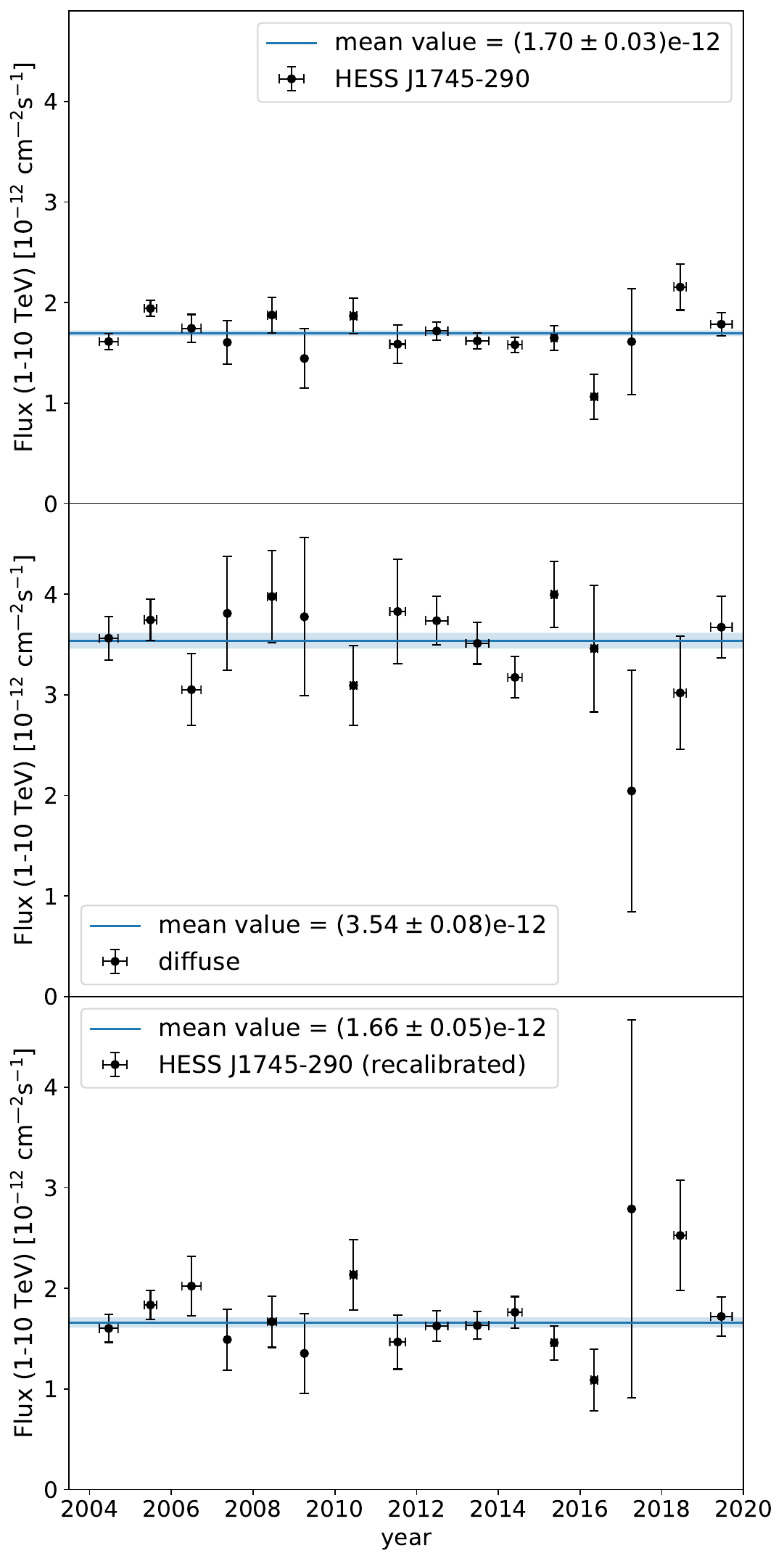}
    \caption{Light curves of \hessGC and DE. Black markers indicate fitted fluxes for each calendar year. Vertical error bars show the statistical error on the fitted flux for that year. Horizontal error bars indicate the span of the observations within each year. The blue line represents the statistical mean of the light curve, with flux uncertainties accounted for, and the band represents the uncertainty on this value. Top: Raw \hessGC light curve. Middle: DE light curve. Bottom: Recalibrated \hessGC light curve.}
    \label{fig:LC}
\end{figure}

\begin{table*}
    \centering
    \caption{Tested evolution scenarios with corresponding $\chi^2$/degrees of freedom.}
    \begin{tabular}{c|cc|cc}
    \toprule
    \toprule
         \textbf{Model}   &   \multicolumn{2}{c|}{\textbf{Raw flux}} &   \multicolumn{2}{c}{\textbf{Recalibrated flux}} \\         
         Parameters   & Fitted values &  $\chi^2$/dof & Fitted values &  $\chi^2$/dof\\
          \midrule
          \makecell{Constant\\ Mean flux [$\times 10^{-12}\,\mathrm{cm^{-2}s^{-1}}$]}     & \makecell{$1.70\pm0.03$ }& 30.4/15  & \makecell{$1.66\pm0.05$ }& 14.9/15\\
         \midrule
         \makecell{Linear\\Gradient [$\times 10^{-14}\,\mathrm{cm^{-2}s^{-1}yr^{-1}} $]} & \makecell{$-0.7\pm 0.9 $}& 28.6/14  & \makecell{$-0.7\pm 1.1 $}& 14.5/14\\
         \midrule
         \makecell{Constant-by-era\\Mean fluxes [$\times 10^{-12}\,\mathrm{cm^{-2}s^{-1}}$]}& \makecell{$1.75\pm0.05$\\$1.63\pm0.07$}& 26.2/14 & \makecell{$1.69\pm0.06$\\$1.63\pm0.09$}& 14.6/14\\
         \bottomrule
    \end{tabular}
    \tablefoot{%
    Each model is described in Section \ref{section42}. 
    All models are compatible with our data, but nonconstant hypotheses are not significantly preferred by our data $($uncertainties on the gradient in the linear model and on both constants in the constant-by-era model make them compatible with a constant solution$)$.
    }
    \label{tab:chi2}
\end{table*}
The light curves presented in Figure \ref{fig:LC} (top and middle panels) show the evolution of the flux of \hessGC and the DE, 
integrated between 1 and 10\,TeV and sampled annually from 2004 and 2019.
The error bars represent statistical uncertainties derived from the fitting of the spectral amplitudes of both components (\hessGC and the DE), while all other parameters are fixed. 
We see large variations in the flux uncertainties, reflecting the large differences in observation time from year to year. 
The two light curves are broadly distributed around a constrained mean value of $1.7 \pm 0.03 \times 10^{-12}$ cm$^{-2}$ s$^{-1}$ for \hessGC (raw) and $3.54 \pm 0.08 \times  10^{-12}$ cm$^{-2}$ s$^{-1}$ for the DE. The difference in the uncertainty values of the two emissions is a logical consequence of the widespread nature of the DE, which implies greater uncertainty in the measurement of its flux. The light curve for \hessGC is broadly distributed around its mean value but shows some deviations beyond the statistical error, notably in the years 2005, 2016, and 2018.
Although the DE light curve is globally consistent with a constant-flux model, year-to-year deviations exceeding the statistical uncertainties are observed. As the source is not expected to vary on yearly timescales, these deviations are likely due to residual systematic uncertainties, which need to be taken into account.
We note that the flux for the year 2017 is not constraining for this dataset due to its low number of observation hours. We show the recalibrated light curve of \hessGC, using the procedure described in Section 4.1, at the bottom of Figure \ref{fig:LC}.
The errors are wider and the mean value is now estimated at $1.66 \pm 0.05 \times 10^{-12}$ cm$^{-2}$ s$^{-1}$.
Whereas the raw light curve of \hessGC shows possible deviations from the mean value, particularly in 2005,
the recalibrated light curve is fully consistent with a constant flux (see reduced-$\chi^2$ values in Table \ref{tab:chi2}).
From the light curve shown in the bottom panel of Figure \ref{fig:LC}, we conclude that a model in which the flux is constant is consistent with the observations (see the ``Constant'' model in Table \ref{tab:chi2}).
None of the time-varying models that we tested provides a better description of the data. 
Indeed, a linear-varying model (a simple linear evolution of the flux between 2004 and 2019; ``Linear'' in Table \ref{tab:chi2}) did not improve the quality of the fit significantly. 
To test for a possible discrepancy between data obtained before and after the passage of the G2 object, as observed by XMM-Newton with the bright X-ray flare rate in \citet{2015MNRAS.454.1525P} (see discussion in section \ref{section61}), we attempted to fit fluxes before and after the end of 2013 separately, using a constant value for each period (``Constant-by-era'' in Table \ref{tab:chi2});
this model does not improve the fit over the constant model (see Table \ref{tab:chi2} for numerical results). 

Finally, we investigated the possibility of energy-dependent time variability.
Some models of the VHE emission from \sgra suggest that a time-varying hadronic process could result in a variable emission with energy-dependent timescales \citep{2011MNRAS.410.1521B}.
We investigated this in two ways. First, we performed a time-resolved 3D analysis in which the spectral indices and the normalizations of the DE and \hessGC models were left free to vary. The cut-off energy of \hessGC was kept constant because it is strongly correlated with the spectral index. 
Applying the same analysis did not reveal any significant variation of any parameters over the years (see appendix A).
Second, we performed the previous time-resolved analysis (with only normalizations free to vary) on different energy ranges. 
The resulting light curves did not show any variation, and neither did several other tests, such as the hardness-ratio test, doubling timescales \citep{Roy_2023}, and the fractional variability \citep{2019Galax...7...62S}. 

\section{Sensitivity to time variations}\label{section5}
\subsection{General method}\label{section51}
Our analysis indicates that the flux of \hessGC shows no significant variability. Here, we assess whether our dataset is sensitive to several theoretical scenarios that produce a time-dependent signal.
A sensitivity study of this kind allows us to quantitatively rule out theoretical scenarios of source variability. 
For a variable source, this method can provide insight into the intrinsic behavior driving its variability.

To test the detectability of a given scenario, we simulated data cubes based on a theoretical source model (one for each year, depending on the time evolution scenario) using the simulation feature of Gammapy. 
We generated model predictions by convolving the models with the IRFs, and the 
resulting  predicted counts were sampled assuming Poisson statistics to produce simulated observations that include noise. 
By repeating this sampling and model-fitting procedure several hundred times and averaging, we tested whether the measured result was compatible with the model within the uncertainties of the fitted model parameters. 
 
In the following subsections, we describe tests of two types of theoretical variation: a one-off variation over a year relative to a constant flux value, and a linear negative variation over 16 years, as predicted by the model of \citet{2013JPhG...40f5201G}. 

\subsection{Overall sensitivity to yearly variations}\label{section52}
\begin{figure}
    \centering
    \includegraphics[width=0.48\textwidth]{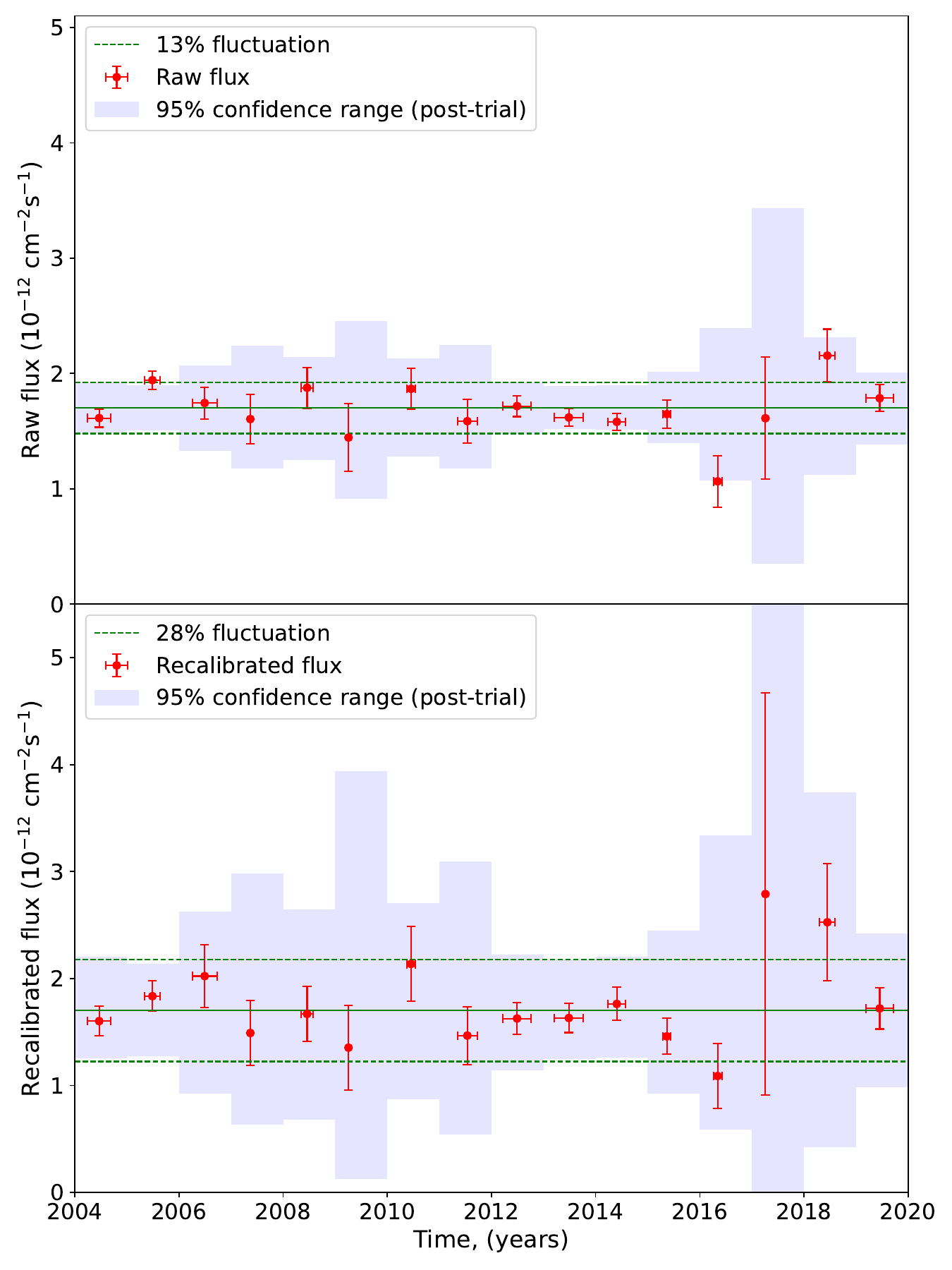}
    \caption{Post-trial 95\% confidence ranges of \hessGC flux per year (blue) when simulations assume a constant flux (solid green line). Top: Raw source measurements. Bottom: Recalibrated flux. The ranges are corrected for the number of trials. Data points from the analysis are shown in red in both panels.}
    \label{fig:simu_constant} 
\end{figure}

The first test uses a constant hypothesis for the flux of \hessGC.
Its aim is to find the range of variation within which we cannot distinguish variation from statistical noise of a constant flux (for a 
95\% confidence interval of sensitivity, corrected for the 16 trials). 
Thus, even without evidence for the variation of \hessGC, we can estimate the minimum variation rate that we can expect to detect, given our data and instrument. 
In practice, we used one dataset for each year, all with the same model;we simulated the data cubes 500 times, and then fit our previous 3D model onto the simulated maps, with only the normalizations free to vary.
We obtained a Gaussian distribution of simulated \hessGC fluxes for each year, which we used to extract the  
95\% confidence range of fluctuations of the multi-year flux (then corrected for the number of trials). 
Figure \ref{fig:simu_constant} displays these results.
The smallest of these ranges provides a lower bound on the sensitivity to deviations from a constant solution over this period. From our study, this minimal sensitivity is estimated to be around 27\% (see Figure \ref{fig:simu_constant}). 
This indicates that, given the time span of the \hess observations of \hessGC, we cannot expect to detect any significant flux variation smaller than 27\% in a given year, even during the most sensitive year.

\subsection{Sensitivity to particle injection from century-old \sgra flares}\label{section53}
Here we test  
an alternative scenario: a steady phenomenon as the origin of a linear temporal variation of the flux of \hessGC.
Motivated by the models of \citet{2011ApJ...726...60C} and \citet{2013JPhG...40f5201G} (see the discussion in \ref{section63}), we tested the hypothesis that the TeV flux of \hessGC decreases linearly due to the evolution of a particle population injected during the massive, century-old flares of \sgra. 
To test the detectability of a linear temporal variation of \hessGC's flux, we simulated light curves to determine above which slope they display a significant linear variation of the $\gamma$-ray flux compared to a constant flux scenario.
As before, we simulated the yearly datasets assuming different time-varying models to generate approximately 10,000 simulated light curves for each scenario.
We tested linear decreases in flux of 0, 10, 20, 30, or 40\% over 16 years.
For each light curve, we tested the compatibility of a linear scenario and a constant scenario using a $\chi^2$ test. If the difference in $\chi^2$ exceeded $(> 3\sigma)$,  the linear model provides a significantly better fit than the constant model, which we interpreted as a significantly detected variation.
At the statistical level, we assessed how often a specific theoretical variation scenario produces a light curve with significant variation.
We assumed that if a scenario produced a significantly observed variation in more than 68\% of the simulations, it would have caused a significant variation within the \hess dataset.

\begin{figure}
    \centering
    \includegraphics[width=0.51\textwidth]{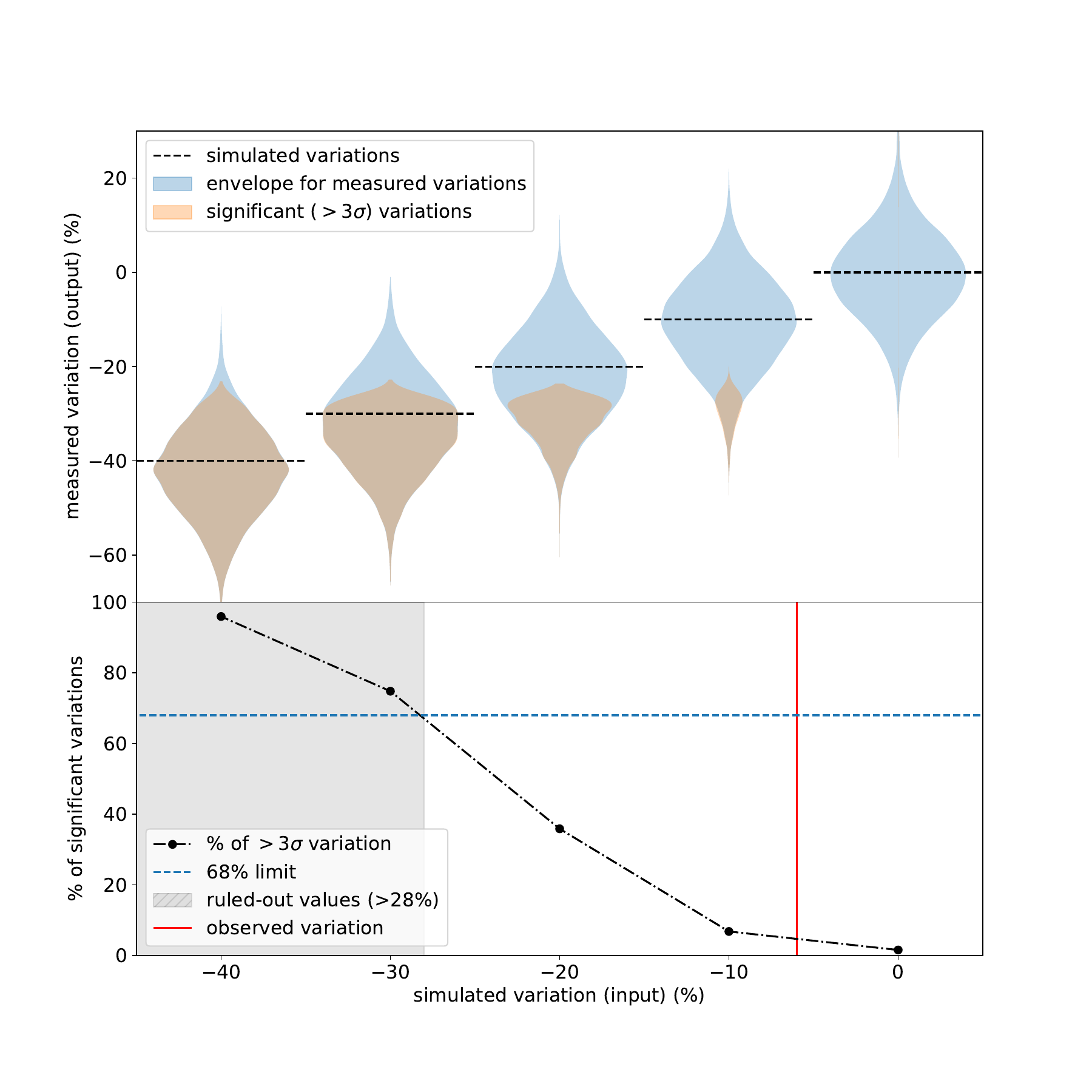}
    \caption{Top: Results of the simulated linearly varying source. Each violin plot shows the distribution of detectable variations over 16 years for each simulated intrinsic variation (from 0 to -40\%; a dashed black line indicates that the mean of this distribution equals the simulated value). The full distribution is shown in blue, and the distribution of detected variations that are significantly preferred to a constant solution is shown in orange. Bottom: Percentage of simulated light curves exhibiting a significant linear variation over 16 years compared to a constant flux model, as a function of simulated source flux linear variation. This percentage increases with the amplitude of the variation, reaching 68\% (dashed blue line) for a simulated linear variation of -28\% (corresponding to a flux decrease of -1.75\% per year over 16 years). The shaded region represents values of simulated linear variations that induce a detectable variation (at the 3$\sigma$ confidence level) with a probability greater than 68\%. The best-fit linear variation actually observed on the data (though not significantly) is shown by the red line}
    \label{fig:sensitivity_linear}
\end{figure}

Figure \ref{fig:sensitivity_linear} illustrates the results. 
The top panel shows the distribution of fitted linear variations for the five theoretical intrinsic linear flux decreases. 
We distinguish between ``observed'' light curves for which the linear model is preferred and those in which it is not.
On average, the simulated variations follow the theoretical value (blue distributions, centered on the dashed black lines).
However, if one considers only the detected variations, we see a bias toward higher values (the orange distributions are not centered on the dashed lines). 
This is due to a threshold below which \hess cannot detect a 16-year linear variation. 
The bias decreases for larger theoretical variations (roughly over 40\%).
Hence, this confirms that, given the sensitivity of the \hess dataset, any detected variability 
is associated with a real variation in the source flux. 
The bottom panel of figure \ref{fig:sensitivity_linear} quantifies the percentage of detected variations among the trials, as illustrated in the top panel. It shows, given a theoretical variation of the \hessGC source, the percentage of light curves that exhibit a significant linear decrease. The detection limit corresponds to a decrease of -28\% (corresponding to -1.75\% of the flux decrease per year), which rules out any larger flux variation over the last 16 years. This analysis confirms that testing a variation hypothesis over the entire period, rather than a single year, enables the detection of much smaller year-to-year variations.
In the next section, we discuss the implications of these results for various emission scenarios. 

\section{Discussion}\label{section6}
Particles accelerated in the inner accretion flow are expected to escape and propagate in the ISM surrounding \sgra, particularly within the 1.5-pc radius CND. Flux variations in the $\gamma$-ray emission are therefore modulated by the $\gamma$-ray production timescale, which depends on particle transport and matter distribution. Variability of \hessGC could therefore occur on timescales of months and years following a possible change in particle injection. 

\subsection{Yearly variation as a result of a transient event}\label{section61}
The sudden change in particle injection and acceleration in the accretion flow could result from transient phenomena near \sgra, which would imply a rapid modification of the SMBH accretion rate. For instance, the model of \cite{2015MNRAS.450..277C} shows how objects falling into the accretion flow can alter the luminosity of \sgra. In 2013-2014, an object known as G2 passed its pericenter around \sgra \citep{2014ApJ...796L...8W}. The passage of this G2 object was quickly proposed as a potential trigger for such a phenomenon,  as observed by \cite{2021ApJ...920L...7M}, and has motivated searches for visible counterparts of the SMBH activity.  Notably, Chandra and XMM-Newton have detected a change in the bright X-ray flare rate since 2013  \citep{2015MNRAS.454.1525P, 2017AetA...604A..85M, 2020AetA...636A..25M}. Although X-ray flares continue at a regular pace, the proportion of brighter flares has increased since that time. Although the overall X-ray flare has remained constant over two decades, \citep{2020AetA...636A..25M} find that the rate of the brightest X-ray flares increased by a factor of 2.5 after mid-2014, suggesting a change in the activity of \sgra. However, the link between this evolution of \sgra's X-ray activity and the G2 passage is doubtful, since the G2 object appears to be only one of many similar objects orbiting close to \sgra around that time, and thus its influence on the SMBH was likely much smaller than initially suspected \citep{2019ApJ...884..148B}. However, analyses of X-ray flares monitored by Swift \citep{2022MNRAS.510.2851A} have also shown a higher activity of \sgra in 2006-2007 and 2017-2019 compared to 2008-2012, although Swift could not monitor the \GC between 2013 and 2016 due to a nearby transient X-ray source.
Assuming that matter fell towards \sgra and altered the accretion flow, and assuming that the particle acceleration rate is proportional to the accretion rate, the TeV emission of the \GC could be affected on comparable timescales, provided the radiation timescale is small.
We tested the presence of any such significant variation in TeV luminosity before and after 2013 on an annual baseline, but we did not see any indication that the TeV emission seen by \hess was altered around or after this year at this magnitude. However, we note that the overall fluence of the bright flares they considered (a few $10^{40}~\mathrm{erg}$ in a year) is small compared to the total fluence of \sgra over a comparable period. 
Furthermore, the TeV emission may include contributions from both the inner accretion flow of \sgra and the surrounding medium due to escaping particles, implying that even sudden and large variations in the accretion rate, directly traced by X-ray flares, could be significantly diluted and contribute only modestly to the measured TeV flux.

\subsection{Long-term evolution of the accretion rate}\label{section62}
We now consider the possibility that a long-term change in the injection and acceleration regime in the accretion disk of \sgra could have occurred over the last decades, gradually affecting the amplitude of the TeV emission over the last 16 years. 
Testing these aspects is particularly interesting because while small short-term variations are limited by the \hess sensitivity, extending the monitoring time of the \GC over several years allows stronger constraints on long-term variations.
Recently, a study of sub-millimeter data involving multiple instruments and their surveys of \sgra over 20 years \citep{2021ApJ...920L...7M}, found the variability of the mean flux to be approximately  $\pm$10\% of the 14-year average (2005-2019). More recently, a 20\% increase was measured in 2019 compared to the previous epoch (2015-2017) \citep{2020AetA...638A...2G, 2022ApJ...931....7B, 2023ApJ...954L..33W}. In this study, the averaged submillimeter fluxes are associated with a population of thermal electrons and are expected to vary together with the mean accretion rate of \sgra \citep{2014ARAetA..52..529Y}.
Although we cannot yet place a constraining limit on this level of variation, our analysis excludes the possibility of a specific increase of flux in 2019 greater than 27\%.  

\subsection{Remnants of major past bursts of \sgra}\label{section63}
There are also indications that \sgra itself was considerably more active in the recent past. The evolution of a large population of accelerated particles injected during the massive, century-old flares of \sgra \citep{2018AetA...610A..34C} could also cause a long-term variation of the TeV emission, as suggested by \citet{2011ApJ...726...60C} and \citet{2013JPhG...40f5201G}. X-ray surveys of nearby clouds in the central molecular zone show temporal brightness variations, particularly in the Fe K$\alpha$ line, which have been explained as echoes of several moderately distant and very bright events \citep{2013AetA...558A..32C,2017MNRAS.465...45C,2018AetA...610A..34C}. This increase was interpreted as a considerable increase in X-ray activity at the GC ($10^5$ times the current luminosity) about one to two hundred years ago. 
Several authors have proposed that this implies that a large number of cosmic rays were accelerated at that time and have since been diffusing away from \sgra into the ISM \citep{2011ApJ...726...60C, 2013JPhG...40f5201G}. In this case, the resulting VHE $\gamma$-ray emission should decrease with time on a timescale dependent on the diffusion model, but likely on the order of decades or longer.  As the initially accelerated CRs diffuse away into the ISM, their density near \sgra decreases as $\sim t^{-3/2}$ after a certain amount of time \footnote{The TeV flux should initially rise as the injected protons propagate through the ISM to cover the dense clouds and then decrease with the local density of CR}.
If most of the VHE $\gamma$ rays result from the interaction of the CRs with the circumnuclear ring (as suggested by \cite{2012ApJ...753...41L}), then the resulting TeV emission should be roughly proportional to the CR density in the central few parsecs around the SMBH, following  $F(t) = F_0t^{-3/2}$, where $t$ is the time elapsed since the powerful flare, and $F_0$ is the reference flux. The ratio between the flux at times $t_1 < t_2$  is then $\frac{F(t_2)}{F(t_1)} = \left(1 + \frac{\Delta t}{t_1}\right)^{-3/2}$.
According to \cite{2017MNRAS.465...45C} and \cite{2018AetA...610A..34C},  
a flaring event approximately 100-200 years ago can explain the X-ray echoes detected across the CMZ.
For our window of observation, $\Delta t = 16$ years, and assuming a powerful injection 100\,years ago (from 2004), the relative decrease in TeV flux over the 16 years should be on the order of 20\%. If we instead consider a 200-year-old flare, this relative decrease drops to 11\%. According to our results, \hess is not yet sensitive to such a resulting decrease in TeV flux over the last 16-years.

\section{Conclusion}
In this work, we analyze the data collected by \hess on \hessGC between 2004 and 2019, making it the longest data set on this region at TeV energies.
This work represents the first application of spectro-morphological analysis to H.E.S.S. observations of the Galactic center region. 
The updated spectrum of \hessGC is compatible with the spectrum previously published in \citep{HESS2016}.
We searched for variability in the flux of \hessGC flux on a yearly basis.
We considered the variability of the central source with respect to the DE by recalibrating the yearly flux of \hessGC with the DE flux. 
We adopted this procedure to mitigate any residual time dependent systematic effects which could  affect a 16-year-long survey. 

As a result, this new study of the GC with \hess finds that the central source \hessGC appears to be stable during this period, corroborating the recent findings of MAGIC \citep{2017AetA...601A..33A} and VERITAS \citep{2021ApJ...913..115A}. A sensitivity study shows that our observations rule out year-to-year luminosity variations larger than 27\% and linear decays larger than 28\%, over the whole time range. 
Our constraints on 
a hypothetical variability
of the central source show that the sensitivity of the instrument to a flux variation (i.e., its capability to detect a flux variation given a certain level of systematics) is not sufficient to rule out a number of time-evolution scenarios for the source. 
Hence, future monitoring of the Galactic center by CTA, with its enhanced sensitivity and improved control of systematics, should help to better constrain \hessGC's temporal behavior.

\bibliographystyle{aa} 
\bibliography{bibtex}

\begin{appendix}

\section{Yearly energy fluxes of \hessGC}

\begin{table}[!ht]
\caption{Recalibrated yearly energy fluxes of \hessGC.}
\label{tab:yearly_flux}
\centering
\begin{tabular}{cc}
\hline\hline
Year & Flux (10$^{-12}$ erg cm$^{-2}$ s$^{-1}$) \\
\hline
2004 & 1.60 $\pm$ 0.14 \\
2005 & 1.84 $\pm$ 0.14 \\
2006 & 2.02 $\pm$ 0.30 \\
2007 & 1.49 $\pm$ 0.30 \\
2008 & 1.67 $\pm$ 0.26 \\
2009 & 1.35 $\pm$ 0.40 \\
2010 & 2.14 $\pm$ 0.35 \\
2011 & 1.47 $\pm$ 0.27 \\
2012 & 1.63 $\pm$ 0.15 \\
2013 & 1.63 $\pm$ 0.14 \\
2014 & 1.76 $\pm$ 0.16 \\
2015 & 1.46 $\pm$ 0.17 \\
2016 & 1.09 $\pm$ 0.30 \\
2017 & 2.79 $\pm$ 1.88 \\
2018 & 2.53 $\pm$ 0.55 \\
2019 & 1.72 $\pm$ 0.19 \\
\hline
\end{tabular}
\tablefoot{%
The yearly energy fluxes are recalibrated according to the method described in Sect.~4.1. Fluxes are given in units of 10$^{-12}$ erg cm$^{-2}$ s$^{-1}$. Uncertainties correspond to 1$\sigma$ statistical errors.
}
\end{table}

\section{Spectral index variation}

\begin{figure}[h]
    \centering
    \includegraphics[width=0.51\textwidth]{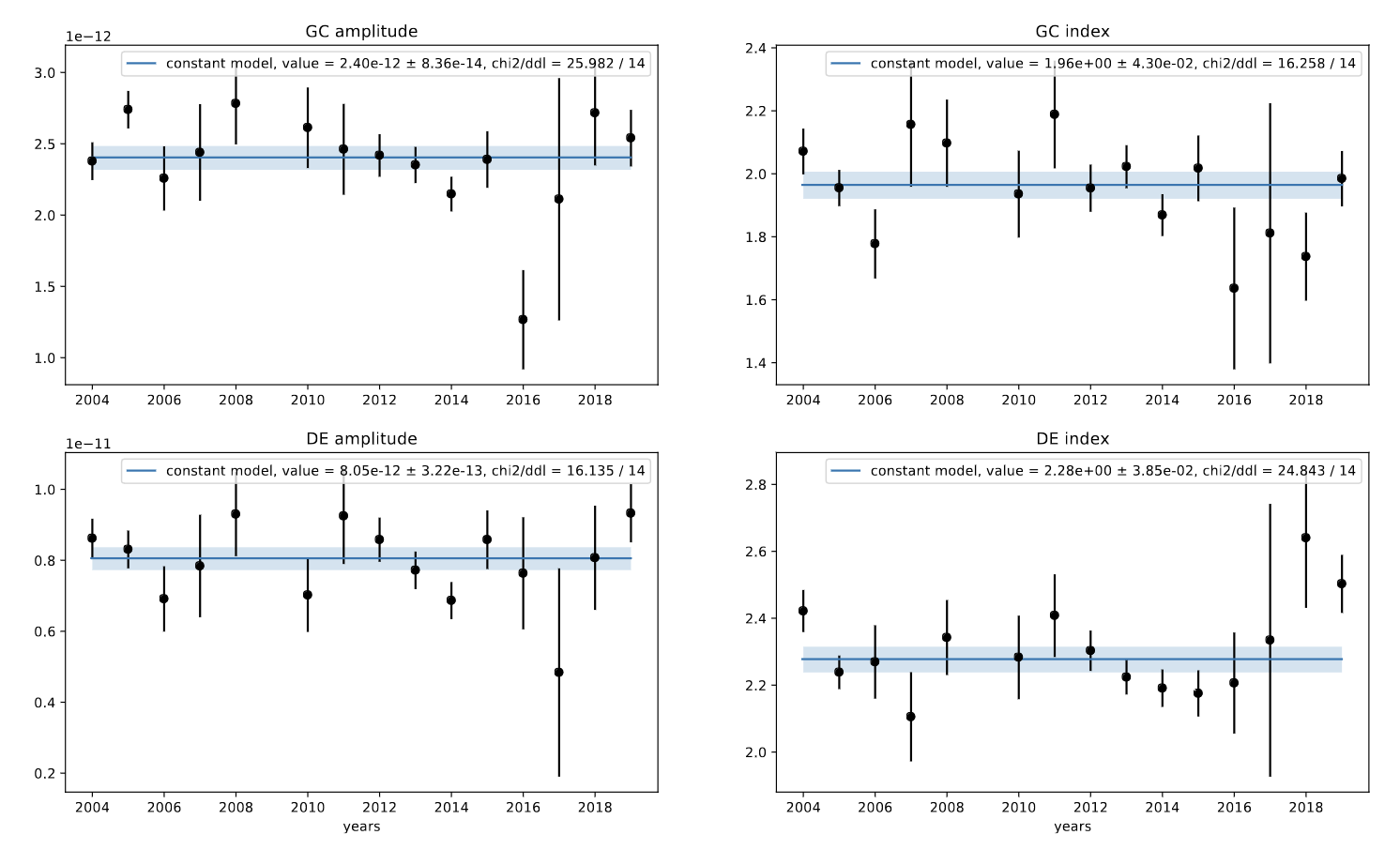}
    \caption{Evolution of the spectral index and normalization over time when left free to vary. Top: evolution of (left) the amplitude normalization and (right) spectral index of \hessGC with time. Bottom: evolution of (left) the amplitude normalization and (right) spectral index of the DE with time. In each panel, the blue line shows the best-fit constant value with its associated uncertainty. The corresponding $\chi^2/\mathrm{d.o.f.}$ values are indicated, showing that in all cases the parameters are compatible with being constant.}
    \label{fig:index_variation}
\end{figure}

We investigated the possibility of energy-dependent time variability of \hessGC. We performed a time-resolved 3D analysis where spectral indices as well as normalizations of the DE and \hessGC are left free to vary during the fit procedure, whereas the cut-off energy of \hessGC spectrum is kept constant. We can see on figure \ref{fig:index_variation}, that it did not result in any significant variation. 

\end{appendix}

\end{document}